# Enhanced low field magnetoresistance of $Fe_3O_4$ nano-sphere compact

P.Y. Song

*Department of Physics, Peking University, Beijing, 100871, China*

J.F. Wang

*Department of Chemistry, Tsinghua University, Beijing, 100084, China and*

*Department of Physics, Wuhan University, Wuhan, 430072, China*

C.P. Chen [a]

*Department of Physics, Peking University, Beijing, 100871, China*

H. Deng and Y.D. Li

*Department of Chemistry, Tsinghua University, Beijing, 100084, China*

## Abstract

Unusually large low field magetoresistance (LFMR), ~ 10 %, at 300 K has been observed with the sample of mono-dispersed $Fe_3O_4$ magnetite nanospheres, ~ 200 nm, compactly cold-pressed and sintered at 800 °C. A detailed analysis on the transport and magnetic measurements indicates that the electron conduction is dominated by the spin-dependent scattering or tunneling at the grain boundaries. At low temperature, 140 K and 100 K near the Verwey transition, ~ 115 K, the LFMR (below 2 kOe) does not show any sign of dependence on the transition, and does not follow the variation of magnetization to reach the saturation region either. On the other hand, at 300 K, the MR saturates fast with the magnetization below 2 kOe. This temperature dependent property in LFMR is very likely attributed to the scattering or tunneling of the conduction electron passing through the grain boundary layer with spin disordered state.

---

[a] e-mail : cpchen@pku.edu.cn,   Phone : +86-10-62751751, Fax : 86-10-62751615

Introduction

In recent years, the extrinsic magnetoresistance (MR) in magnetic oxides, such as $La_{0.3}Sr_{0.7}MnO$, $CrO_2$ and magnetite, etc., have been intensively studied due to the application potential at a possibly lower cost in comparison with the giant magnetoresistance (GMR) sensors.[1] The extrinsic MR is arising mostly from the effect of spin-dependent interface scattering or tunneling. These studies have been carried out in a hope to optimize the devices to have a room temperature MR effect competitive with the GMR mutlilayers. The ferrimagnetic magnetite, $Fe_3O_4$, with a Curie temperature at 858 K, is a very good candidate to show the required properties. It has an exceptionally high conductivity among the ferrites. The resistivity at 300 K is on the order of $10^{-2}$ $\Omega \cdot cm$,[2] which is an order of magnitude higher than the minimum metallic resistivity. In addition, the band-structure calculations have predicted that $Fe_3O_4$ is a half-metal with only one spin-subband at the Fermi level,[3,4] Experimentally, the spin-resolved photoemission on the $Fe_3O_4$ epitaxial thin film indeed indicates a spin polarization around - 80 % at room temperature.[5] MR value as high as 13 % at 300 K has been reported in the (cobalt/alumina/iron oxide) junction.[6] This large MR has been ascribed to the presence of $Fe_{3-x}O_4$, which is close to a half-metallic magnetite. The half metallic nature of $Fe_3O_4$ makes it especially promising for the application in the spin-electronic devices.

Single crystal $Fe_3O_4$ has a Verwey transition at about 117 K. It shows a non-metallic transport behavior between the Verwey transition temperature and 320 K. The resistivity jumps sharply by nearly two orders of magnitude as the temperature decreases crossing the Verwey temperature.[7] Various forms of magnetite have been studied, such as single crystal,[2,6] polycrystal,[2] epitaxial and polycrystal films,[8-10] magnetic tunneling junctions,[11] and granular systems.[12,13] However, only a few progresses have been achieved, especially with the LFMR, which is of more value for application.

There have been numerous attempts to fabricate polycrystalline $Fe_3O_4$ samples with high MR value in a low applied field at 300 K. Although MR more than 20 % has been achieved in a high applied magnetic field, ~ 40 T, at about 200 K,[13] it is only a few percent in the applied field of the order of 10 kOe at room temperature[8-13] with one exceptional case using the nano-contact.[14] In the present article, we report enhanced LFMR with the compact of mono-dispersed $Fe_3O_4$ magnetite nanospheres, ~ 200 nm. The magnitude exceeds 10 % in the applied field roughly equal to 2 kOe at room temperature.

Sample preparation and characterization

The mono-dispersed nanopowder of magnetite was synthesized using the solvothermal reduction method described in detail in the previous report.[15] In brief, certain amount of $FeCl_3 \cdot 6H_2O$ is dissolved into ethylene glycol, then a little amount of NaAc and polyethylene glycol is added into the solution. Afterwards, the mixture is heated and maintained at 200 °C for 8 hours. The $Fe_3O_4$ nano-particles would then be produced by the reduction reaction. The average size of each grain is about 200 nm estimated from the SEM images, see Fig. 1. The nanopowder was then cold-pressed into a disk-shaped sample under a pressure of 4 MPa and sintered subsequently for 3 hours at T ~ 800 °C in the Ar atmosphere. The peaks in the x-ray diffraction pattern ($CuK_\alpha$) of the synthesized nanopowder, shown in Fig. 2, can all be indexed in the $Fe_3O_4$ spinel structure ( JCPDS No. 79-0417). By the Rietveld analysis of the XRD data, the lattice parameter is calculated as $a = 0.8392$ nm in comparison with the listed value of 0.8394 nm (JCPDS 79-0417). The XRD analysis indicates that the sample is of single phase.

Measurements and analysis

The transport properties were measured using Quantum Design physical property measurement system (PPMS) and the magnetization, using magnetic property measurement system (MPMS). Figure 3 shows the zero-field-cooling (ZFC) and the field-cooling (FC) temperature dependent magnetization, recorded on warming in a magnetic field of 90 Oe. For FC measurement, the sample was cooled in the field of 20 kOe before warming up for data collection. The sample shows a clear Verwey transition at about 115 K, determined from the maximum slope of ZFC magnetization. It marks a structural transition from a cubic high-temperature phase to a monoclinic low-temperature one.

The temperature dependent resistance (R-T) was measured by a 4-probe technique, using silver paste as contacts. The curve in Fig. 4 shows a non-metallic behavior from 50 K to 300 K. At 300 K in the absence of an applied field, the resistivity is estimated as 0.5 $\Omega \cdot m$. It is similar to that observed with a compacted magnetite nano-powder and about 3 orders of magnitude larger than those observed with a bulk single crystal or thin films.[2] The much larger resistivity indicates that most of the resistance arises from the scattering or tunneling at the interfaces between the nanoparticles. At low temperature, the R-T curve appears to be smooth in Fig. 4 across the Verwey temperature. It indicates that the extrinsic effect is large enough to obscure the intrinsic one. Therefore, it is reasonable to estimate the averaged contribution of the contact resistance per interface at 300 K from the resistivity. The estimation gives $\rho_{300K}/D \sim 2.5$ M$\Omega$, where D ~ 200 nm, is the size of particle. This contact resistance is much larger than the quantum limit of a metallic contact, $h/2e^2 = 12.9$ k$\Omega$. Thus the electron transport exhibits non-metallic nature. In the inset of Fig. 4, the resistance is plotted on a logarithmic scale versus $1/T^{1/2}$. The data points apparently fall on a straight line. The solid line is drawn for the guide of eye sight. This behavior, with the resistance varying as $\exp(T_0/T)^{1/2}$, can be explained by the hopping transport owing to the granular nature of the sample.[16,17]

The MR measurement at 300 K has been performed using a 4-probe technique by sweeping the applied field from 20 kOe down to – 20 kOe and then back to 20 kOe for a complete cycle. The MR is defined as, $MR = (R_H - R_0)/R_0$, where $R_0$ is the resistance at zero field and $R_H$ is that measured in an applied field, H. With the present sample, the MR plotted in Fig. 5 is negative as expected. Also plotted in the same figure is the field-dependent magnetization. A very interesting point to note is that the MR reaches the saturation region following the field response of the magnetization, M(H), as shown in the figure. In the low field region, -2 kOe < H < 2 kOe, the MR varies fast. It strongly correlates with the variation of M(H) and reaches within 20 % of the value taken in the applied field of 20 kOe. The MR at 300 K in the applied field of 2 kOe is about 9 ~ 12 %. This is a very large effect in comparison with the results reported in the other experiments with similar conditions.[12,13] Also, it is observed that when the magnetic field has swept a complete cycle, the resistance becomes higher than the initial starting value. The arrows in Fig 5 indicate the field-sweeping direction. Similar behavior of irreversibility in MR after a complete cycle of field sweeping has been observed in previous experiments with a $Fe_3O_4$ powder compact[2] or a $Fe_3O_4$ film[18]. However, there is not any elaboration on this behavior in these reports. The difference in resistance between the initial and final values amounts to more than 3 % in MR. If the variation in the resistance reflects the temperature variation of the sample, then the increase in the resistance would imply a drop in temperature by about 2 K. This phenomenon can not be explained by the Joule heating of measuring electric current. .The difference in the resistance, $\Delta R$, between the two irreversible branches in R-T measurement at H > 0 has been calculated. It exhibits a feature of constant shift. With the level of measuring current at I = 10 µA, the corresponding voltage shift estimated by $\Delta V = I \Delta R$ is about 20 µV. This implies that the irreversibility is more likely due to the spin-dependent charge accumulation, perhaps, occurring at the contact to the electric leads. Further experiment is needed in order to understand the underlying mechanism causing this behavior. Nonetheless, we have made a correction on the MR value at H > 0 in Fig. 5 to remove the effect of the above-mentioned irreversibility. The resistance in the upper branch has been subtracted by $\Delta R/2$, while the lower one, added by $\Delta R/2$. The corrected MR is then calculated and presented in Fig. 5 by the solid curves. After the correction, the MR in H > 0 and H < 0 appears to be at the same level. In the saturation region, at H = 20 kOe or – 20 kOe, MR ~ 10.9 %.

The MR has been measured at 140 K and 100 K as well. One of them is above the Verwey transition and the other, below. The results are plotted in Fig. 6. In the high field region, at H = 45 kOe, the saturation magnetization is about 94.1 emu/g at 140 K and 94.7 emu/g at 100 K. They correspond to 3.92 and 3.95 $\mu_B$/formula unit, respectively. It is interesting to observe that the MR at the two different temperatures are almost the same, indicating that the Verwey transition has almost no effect at all on the MR. Furthermore, an interesting anomaly appears with the low temperature MR. Unlike at 300 K, the field dependence of MR at low temperature does not follows the magnetization to reach the saturation at the applied field of 2 kOe. Its magnitude increases significantly as the magnetic field goes up. This suggests that the bulk magnetization is not the major factor to show the observed MR effect. The magnetization curves at T = 140 K and 100 K are presented in Fig. 6 as well. It is apparent that the two curves almost collapse. At the field of 20 kOe, the MR reaches only about 4 %, and at 2 kOe, about 0.8 %. This is much less than that at 300 K. This property is not expected and has not been observed in any of the previously reported experiments. Usually, MR at low temperature is higher than that at high temperature. The present result suggests that, at low temperature, there exists an effect dominating the transport process which is not attributed to the interior magnetization of the $Fe_3O_4$ nanoparticle.

Discussion

It has been studied that the tunneling current has exhibited spin-dependent properties strongly dominated by the interfacial magnetic state of $Fe_3O_4$[19] and by the grain boundary magnetic property of $(Ba_{0.8}Sr_{0.2})_2FeMoO_6$[20]. The experimental evidence in the present work has also supported that the magnetic state of the grain boundary is the major factor to yield the large LFMR at 300 K. It suggests that the magnetic moments near the grain boundary respond to the applied field more closely at 300 K than at low temperature. Consequently, the electron tunnels through the grain boundary more easily at 300 K. It would be interesting to further study the boundary magnetic state in order to understand the nature of the observed MR.

Conclusion

In conclusion, the highly enhanced MR, up to 10 %, in the low field region, 2 kOe, has been observed at 300 K, whereas the MR at low temperature, 140 K and 100 K, is much reduced. The experimental evidence indicates that the spin-dependent scattering or tunneling at the interface between two adjacent nano-grains is the major mechanism to result in the observed MR effect. Although the mechanism behind the phenomenon is not yet fully understood and more research works remain to carry on, it is very likely that the surface disordered spin state in the surface layer dominates the transport process. Hence, the surface magnetic property of the nano-grains may be the crucial factor for a large LFMR effect at room temperature, which is important for the purpose of application.

Figure Captions

Fig. 1　SEM image of the $Fe_3O_4$ nano-particles. The averaged size is about 200 nm.

Fig.2.　X-ray diffraction pattern of the magnetite namoparticles. All of the peaks can be indexed to the $Fe_3O_4$ ( JCPDS No.79-0417) with a spinel structure.

Fig. 3　FC/ZFC magnetization. The field of cooling for the FC measurement was 2 Tesla. The magnetization was measured on warming in a magnetic field of 90 Oe. The Verwey transition is determined as 115 K from the maximum slope of ZFC curve.

Fig. 4　Temperature dependent resistance from 300 K down to 50 K. The R-T curve appears smooth around the Verwey transition, ~ 115 K. The inset is plotted with R in a logrithmic scale versus $1/T^{1/2}$. The solid line is for the eye guide.

Fig.5. Magnetoresistance and field dependent magnetization measurements at 300 K. Both quantities reach the saturation values at the field of about 2 kOe, indicating the strong correlation of the two quantities. The arrows indicate the direction of sweeping field. The solid curve is for the MR corrected for the effect of irreversibility in the field sweeping.

Fig. 6　MR (left Y-axis) and magnetization (right Y-axis) at 140 K and 100 K. The MR values reach about 4 % at the field of 20 kOe. The magnetization curves collapse for the ones taken at 100 K and 140 K.

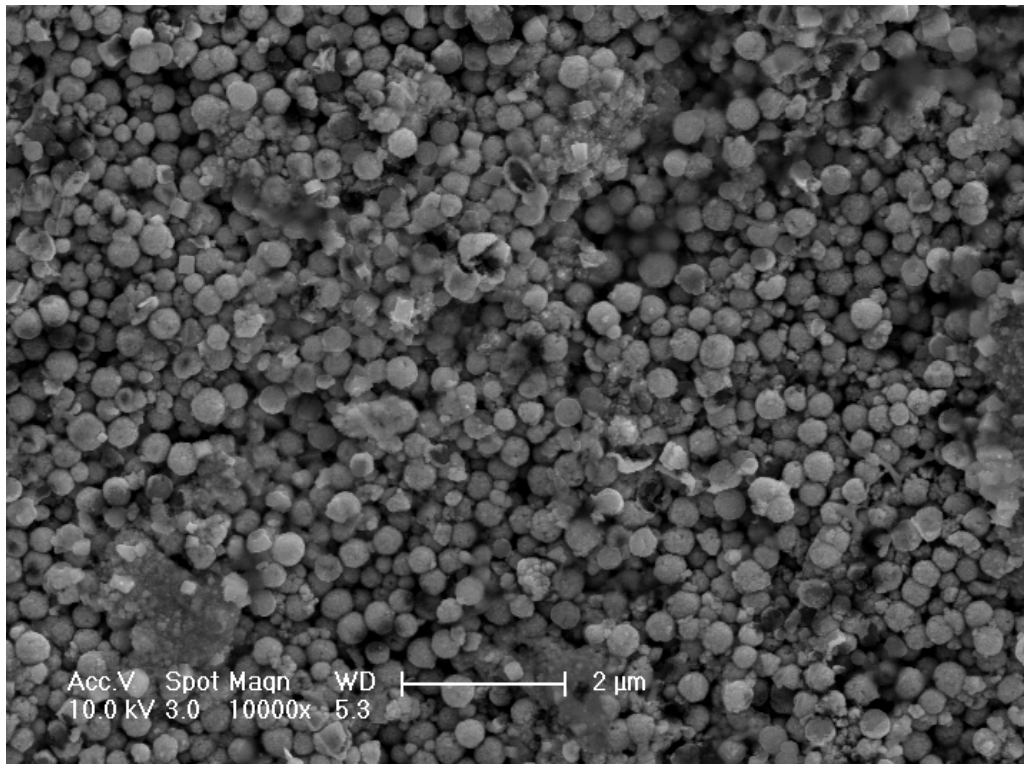

Fig. 1

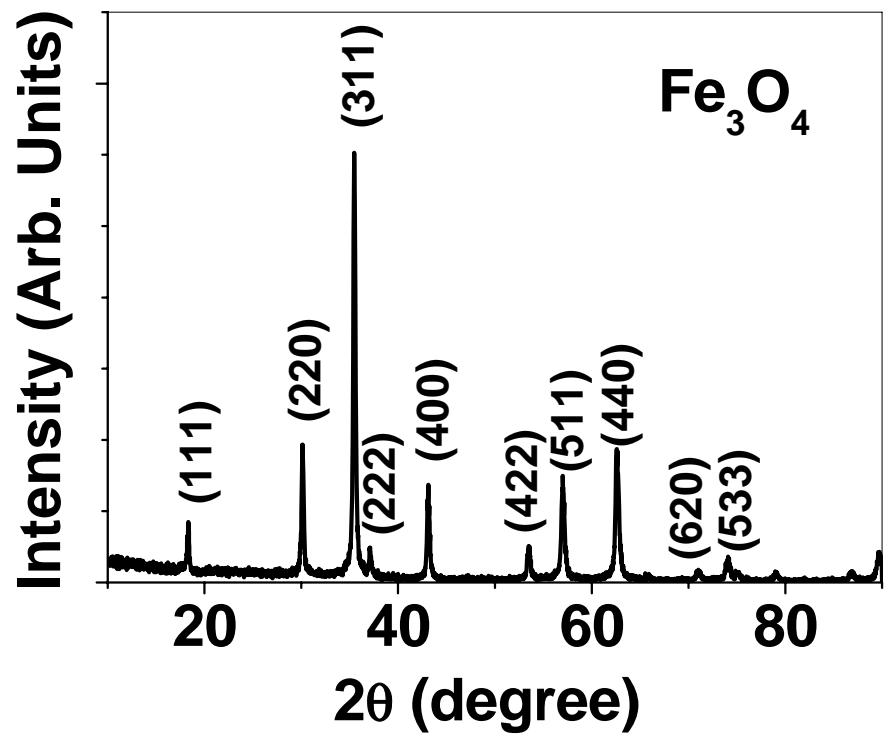

Fig. 2

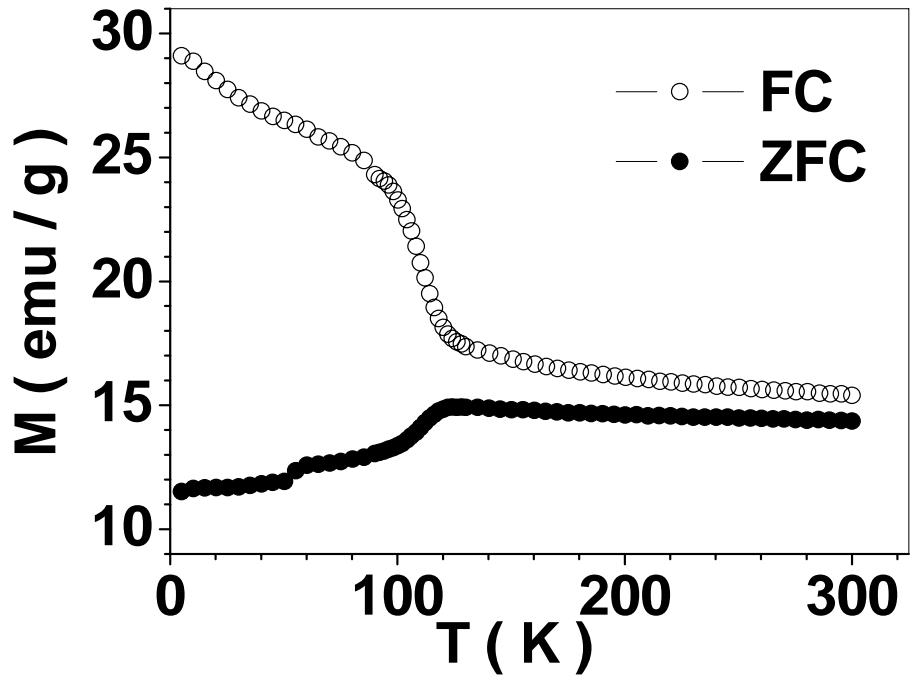

Fig. 3

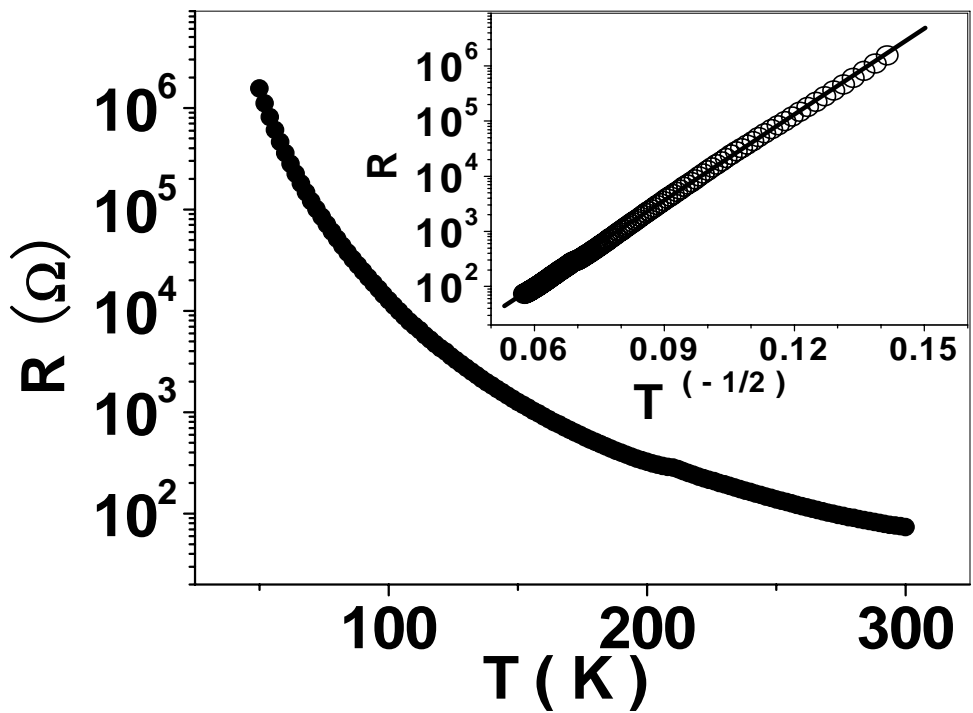

Fig. 4

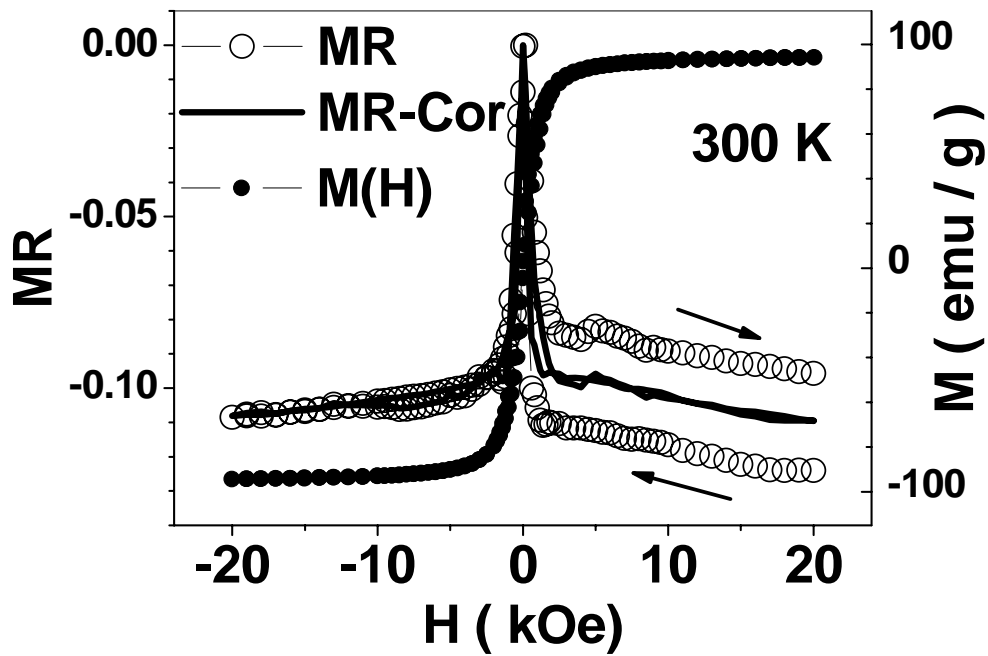

Fig. 5

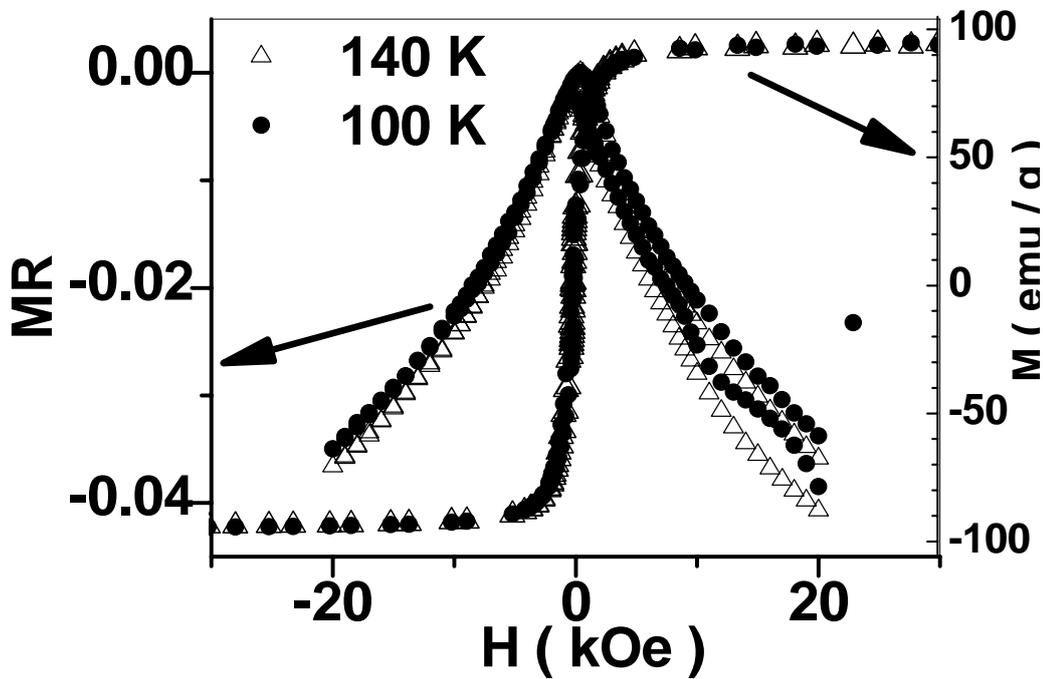

Fig. 6